\documentclass[floatfix,a4paper,aps,prd,twocolumn,preprintnumbers]{revtex4} 
\usepackage{epsfig}
\usepackage{axodraw}
\usepackage{amsmath}
\usepackage{amssymb}
\usepackage{amsfonts}
\usepackage{graphicx}
\usepackage{graphics,subfigure}
\usepackage{axodraw}
\usepackage{float}
\usepackage{booktabs}
\usepackage{color}
\usepackage{geometry}
\usepackage{rotating}
\usepackage{units}
\usepackage{slashed}

\newcommand{\lsim}{\raisebox{-0.13cm}{~\shortstack{$<$ \\[-0.07cm] $\sim$}}~}
\newcommand{\gsim}{\raisebox{-0.13cm}{~\shortstack{$>$ \\[-0.07cm] $\sim$}}~}

\newcommand{\bea}{\begin{eqnarray}}
\newcommand{\eea}{\end{eqnarray}}
\newcommand{\beq}{\begin{equation}}
\newcommand{\eeq}{\end{equation}}
\newcommand{\beqa}{\begin{eqnarray}}
\newcommand{\eeqa}{\end{eqnarray}}
\newcommand{\bit}{\begin{itemize}}
\newcommand{\eit}{\end{itemize}}

\newcommand{\st}{\ensuremath{\tilde{t}_1}}
\newcommand{\neutr}{\ensuremath{\tilde{\chi}_1^0}}
\newcommand{\charg}{\ensuremath{\tilde{\chi}_1^\pm}}
\newcommand{\gev}[1]{\unit[#1]{GeV}}
\newcommand{\mev}[1]{\unit[#1]{MeV}}
\newcommand{\tev}[1]{\unit[#1]{TeV}}
\newcommand{\met}{\ensuremath{\slashed{E}_T}}
\newcommand{\mpt}{\ensuremath{\slashed{p}_T}}
\newcommand{\ifb}[1]{\unit[#1]{fb^{-1}}}

\newbox\charbox
\newbox\slabox
\def\s#1{{      
    \setbox\charbox=\hbox{$#1$}
    \setbox\slabox=\hbox{$/$}
    \dimen\charbox=\ht\slabox
    \advance\dimen\charbox by -\dp\slabox
    \advance\dimen\charbox by -\ht\charbox
    \advance\dimen\charbox by \dp\charbox
    \divide\dimen\charbox by 2
    \raise-\dimen\charbox\hbox to \wd\charbox{\hss/\hss}
    \llap{$#1$}
}}

\begin{document}
 
\title{Light Stop Searches at the LHC in Events with two b--Jets and Missing
  Energy} 

\author{S. Bornhauser}
\email[]{bornhaus@th.physik.uni-bonn.de}
\affiliation{Department of Physics and Astronomy, University of New Mexico,
  Albuquerque, NM 87131, USA} 

\author{M. Drees}
\email[]{drees@th.physik.uni-bonn.de}
\affiliation{BCTP and Physics Institute, University of Bonn, Bonn,
  Germany \hspace*{1cm} and \\ School of Physics, KIAS, Seoul 130--722, Korea}

\author{S. Grab}
\email[]{sgrab@scipp.ucsc.edu}
\affiliation{Santa Cruz Institute for Particle Physics, University of
  California, Santa Cruz, California 95064, USA} 

\author{J.S. Kim}
\email[]{jongsoo.kim@tu-dortmund.de}
\affiliation{Institut f\"ur Physik, Technische Universit\"at Dortmund, D-44221
  Dortmund, Germany}

\begin{abstract}
We propose a new method to discover light top squarks (stops) in the
co--annihilation region at the Large Hadron Collider (LHC). The bino--like
neutralino is the lightest supersymmetric particle (LSP) and the lighter stop
is the next--to--LSP. Such scenarios can be consistent with electroweak
baryogenesis and also with dark matter constraints.  We consider the
production of two stops in association with two $b-$quarks, including pure QCD
as well as mixed electroweak--QCD contributions. The stops decay into a charm
quark and the LSP. For a higgsino--like light chargino the electroweak
contributions can exceed the pure QCD prediction.  We show the size of the
electroweak contributions as a function of the stop mass and present the LHC
discovery reach in the stop--neutralino mass plane.
\end{abstract}

\preprint{DO-TH 10/17}
\preprint{SCIPP 10/19}
\preprint{BONN-TH-2010-08}

\maketitle
 
\section{Introduction}
\label{sec:intro}

The Large Hadron Collider (LHC) is currently collecting data at
$\sqrt{s}=\tev{7}$ and it is assumed that the integrated luminosity will reach
$\ifb{1}$ next year. If low--energy supersymmetry (SUSY) \cite{susy} is
realized, detection of light supersymmetric particles may be around the corner
\cite{lightsusy}.

The scalar top (stop) within the minimal supersymmetric standard model (MSSM)
\cite{susy} is naturally one of the lightest SUSY particles (sparticles). On
the one hand, the large top Yukawa coupling leads to large mixing between
left-- and right--handed stops, reducing the mass of the lightest stop mass
eigenstate.  On the other hand, the large top Yukawa coupling reduces the stop
mass at the electroweak (EW) scale via renormalization group equation (RGE)
running \cite{Ibanez:1984vq}. The lightest stop, $\tilde{t}_1$, is mostly
right--handed, since $m_{\tilde t_R}$ is not increased by $SU(2)$ gaugino
loops, and is more strongly reduced by the top Yukawa interaction
\cite{Ibanez:1984vq}.

A light stop with a mass of $m_{\tilde{t}_1} \lesssim 125$ GeV is vital for
successful EW baryogenesis within the MSSM \cite{Carena:2008vj, Carena:2008rt,
  Li:2008ez, Huet:1995sh, Cirigliano:2006dg, Cirigliano:2009yd}. It allows for
a strong first order phase transition, which prevents the generated baryon
asymmetry of the universe \cite{Nakamura:2010zzi, Komatsu:2010fb} from being
washed out. Furthermore, CP violation is needed in order to generate the
baryon asymmetry \cite{Sakharov:1967dj}. The Standard Model (SM) CP violating
Dirac phase is too small \cite{cpsm}, whereas the MSSM contains additional
CP--violating phases sufficient for EW baryogenesis \cite{Huet:1995sh}.

However, new CP violating phases are strongly constrained by the
non--observation of electric dipole moments ruling out large regions of the
MSSM consistent with EW baryogenesis \cite{susy_edm, Li:2008ez,
  Cirigliano:2009yd, Cirigliano:2006dg}. Of the remaining MSSM parameter
space, scenarios with bino--driven baryogenesis are probably the most
promising ones \cite{Li:2008ez, Cirigliano:2009yd}. Here, CP violation in the
bino--higgsino sector accounts for successful EW baryogenesis. In this
scenario the (stable) lightest supersymmetric particle (LSP) is a bino--like
neutralino and the Higgs mixing parameter $\mu$ needs to be of the order of
the bino mass $M_1$. In addition, all sfermions (beside the light stop) are
quite heavy, in order to suppress electric dipole moments and to fulfill
present bounds on the Higgs mass \cite{Nakamura:2010zzi, Carena:2008vj}.  The
light stop should be predominantly an $SU(2)$ singlet (i.e. ``right--handed'')
in order to suppress stop loop contributions to the electroweak rho parameter
\cite{rho_stop}. As we show below, many of these scenarios fall into the
parameter space that we investigate in this work \footnote{We do not consider
  additional CP--violating phases (compared to the SM one) in this paper. They
  are unimportant for our collider process.}.

In the MSSM, the lightest neutralino, $\tilde{\chi}_1^0$, is a promising dark
matter (DM) candidate if it is the LSP \cite{Ellis:1983ew}. However, large
regions of the MSSM parameter space are disfavored due to a too large relic
density of the $\tilde{\chi}_1^0$ \cite{Baer:2010wm}. The relic density is
determined by the thermally averaged cross section, which includes
annihilation and co--annihilation processes. Other sparticles with masses not
far above that of the LSP can co--annihilate with the neutralino and/or
enhance $t-$ or $u-$channel exchange contributions to the annihilation
processes reducing the DM density in the universe to a level consistent with
cosmological observations \cite{Komatsu:2010fb}.

If the mass splitting between the neutralino and another particle is $\lsim
20\%$, the co--annihilation diagrams are significant \cite{coan}. We consider
such scenarios. We assume a relatively light neutralino LSP, a stop
next--to--lightest supersymmetric particle (NLSP) and a light higgsino--like
chargino $\charg$. All other sparticles are assumed to be heavy. We take
neutralino--stop and stop--chargino mass differences of a few tens of GeV.
These scenarios can be consistent with DM constraints \cite{Balazs:2004bu,
  Cirigliano:2006dg}, {\it i.e.}  the thermal $\tilde{\chi}_1^0$ relic density
is equal to or lies below the observed DM density.

For most of our scenarios the (thermal) $\tilde{\chi}_1^0$ relic abundance
lies below the observed one if $|\mu| \approx |M_1|$ \footnote{We have
  calculated the $\tilde{\chi}_1^0$ relic abundance for various of our
  scenarios with the help of {\tt micrOmegas2.4} \cite{micromegas}.}.  This is
because a non--negligible higgsino component of the $\tilde{\chi}_1^0$ leads
to efficient $\tilde{\chi}_1^0$ annihilation into $W^+W^-$ and $Z^0Z^0$ pairs.
However, there are several options to make scenarios with low thermal
$\tilde{\chi}_1^0$ density phenomenologically viable. The simplest option is
to have an additional DM component, see
e.g. Refs.~\cite{multi_dm}. Furthermore, a non--standard cosmological history
can lead to the right $\tilde{\chi}_1^0$ density \cite{quint}.  Finally, the
$\tilde{\chi}_1^0$ abundance can arise from the decays of metastable species
if their lifetime is so large that they decay out of thermal equilibrium
\cite{nonthermal}. In contrast, dilution of a (too large) thermal
$\tilde{\chi}_1^0$ relic density is more difficult to achieve
\cite{Cirigliano:2006dg}.

Stop pair production might be difficult to detect in the co--annihilation
region at hadron colliders \cite{Balazs:2004bu}. The decays $\st\rightarrow b
\neutr W$, with $b$ ($W$) the bottom quark ($W$ boson), as well as $\st
\rightarrow \tilde \ell \nu_\ell b, \tilde \nu_\ell \ell b$ are kinematically
closed. The loop induced two--body decay $\st\rightarrow c \neutr$ then
competes with tree--level four--body decays like $\st \rightarrow \ell
\nu_\ell b \tilde \chi_1^0$. The latter are strongly phase space suppressed for
small stop neutralino mass splitting, so that the loop induced decay becomes the
dominant decay mode \cite{Hikasa:1987db,abdel}. However, the charm quark, $c$,
is soft, so that the collider signature is given by two soft charm jets and
missing energy.

Tevatron searches for light stops, decaying to charm and $\neutr$, require a
minimum mass gap between the stop and $\neutr$ of at least $\gev{40}$
\cite{Abazov:2008rc, CDFnote9834}. Otherwise the charm jets are too soft to
be seen above the SM backgrounds. As a consequence, the Tevatron is not very
sensitive to the coannihilation region.  At the LHC, the detection of stop
pair production is expected to be even more difficult due to a large hadronic
activity.  However, gluinos can decay into a stop and a top. Due to the
Majorana character of gluinos, same--sign tops can be produced. The resulting
collider signature can be probed at the LHC as long as the gluino mass does
not exceed $\gev{900}$ \cite{Kraml:2005kb}; see also
Refs.~\cite{Martin:2008aw}. Ref.~\cite{Carena:2008mj} therefore investigated
QCD stop pair production in association with a very energetic photon or
jet. The resulting signature is one hard photon or jet recoiling against large
amounts of missing energy.  Their results are independent of the gluino mass
and the discovery reach covers stop masses up to roughly $\gev{240}$ for an
integrated luminosity of $100\,\text{fb}^{-1}$ at $\sqrt{s} = \tev{14}$.

Here we instead consider the production of two stops in association with two
$b$ (anti)quarks. We include pure ${\cal O}(\alpha_S^4)$ QCD diagrams as well
as leading order ${\cal O}(\alpha_s^2 \alpha_W)$ mixed QCD--EW contributions.
The latter are due to diagrams with an on--shell higgsino--like chargino and
substantially increase the total cross section. Information about the
magnitude of the EW contributions can in principle be obtained by measuring
our process {\it and} stop pair production in association with a hard jet
\cite{Carena:2008mj}.  The latter process can be used to determine the stop
mass, which in turn uniquely fixes the pure QCD contribution to our process.
Subtracting this from the measured cross section would yield a determination
of the magnitude of the EW diagrams.

These diagrams are sensitive to the higgsino coupling
$\st-\charg-b$. Therefore, the measurement of our process allows a test of the
respective SUSY coupling relation and thus a test of supersymmetry itself
\cite{Freitas, Allanach:2010pp}. Only a few of such tests have been proposed
so far for the LHC \cite{Freitas, Bornhauser:2009ru, Allanach:2010pp} and none
of them addresses interactions which originate from the superpotential.

The remainder of this article is organized as follows. In
Sect.~\ref{sec:signal_description} we describe our process and present  
the dominant QCD and EW contributions. In Sect.~\ref{sec:lhc_analysis}, we first
discuss the dominant background processes and then basic cuts for a benchmark
scenario before presenting our numerical results. We show the discovery reach
in the neutralino stop mass plane. In Sect.~\ref{sec:charm_tagging}, we
discuss possibilities to further optimize our cuts, which sensitively depend
on the stop neutralino mass splitting. We conclude in Sect.~\ref{sec:summary}.

\section{Stop Pair Production in Association with two b-Jets}
\label{sec:signal_description}

We consider stop pair production in association with two $b-$jets in proton
proton collisions,
\beq
pp\rightarrow \st\st^* b \bar b.
\eeq
We are interested in scenarios where the mass difference between the lightest
stop and neutralino is not larger than a few tens of GeV as discussed in the
introduction. In such scenarios, the stop decays into the lightest neutralino
and a soft charm jet,
\beq
\st\rightarrow \neutr c.
\eeq
Due to the small $\st - \neutr$ mass splitting, much of the time the $c-$jets
will be too soft to be useful \cite{Kraml:2005kb}.  Our hadron collider
signature is therefore large missing energy and two $b-$flavored jets. We
require both hard jets to be tagged as $b-$jets, since this greatly suppresses
SM backgrounds. The dominant QCD contributions are generated via
$gg\rightarrow\st\st^*$ processes, where the $b\bar b$ pair comes from
additional gluon radiation splitting into $b \bar b$. One of the dominant
diagrams is shown in Fig.~\ref{fig:qcd_t1t1bb}. QCD contributions with quarks
and antiquarks in the initial state are subdominant.  For example, for
$m_{\st}=\gev{400}$, quark--antiquark diagrams contribute only about $6\%$ to
the total cross section due to the small $q \bar q$ flux at the relevant
Bjorken$-x$ \cite{Martin:2009iq}; this is smaller than the uncertainty of our
leading order analysis. For smaller stop masses these contributions are even
less important. Therefore, we only consider gluon fusion diagrams; {\it cf.}
Figs. \ref{fig:qcd_t1t1bb} and \ref{fig:ew_t1t1bb}.
\begin{figure}[h]
\includegraphics[width=0.3\textwidth]{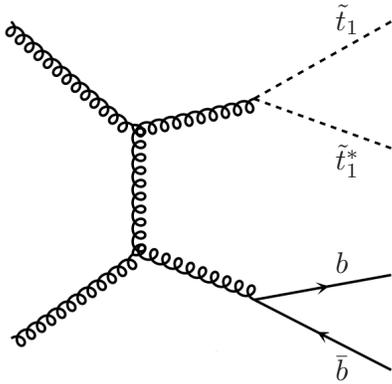}
\caption{Example diagram for QCD stop pair production in association with two
  $b-$jets via gluon fusion.} 
\label{fig:qcd_t1t1bb}
\end{figure}

We also include the leading EW contributions to our process; one of the
corresponding Feynman diagrams is shown in Fig.~\ref{fig:ew_t1t1bb}. Diagrams
with electroweak gauge bosons and Higgs bosons exchange are subdominant and
are not taken into account in our analysis. Again, contributions with a quark
and an antiquark in the initial state are suppressed and we do not consider
them. If an on--shell decay of a chargino into a stop and a $b-$quark is
kinematically possible, the diagram shown in Figure~\ref{fig:ew_t1t1bb} is
effectively only a $2\rightarrow3$ process, because the (on-shell) chargino
will decay into $\tilde t_1 b$; this decay will almost always be allowed if
$\tilde t_1$ is the NLSP. If the chargino mass is not much above the stop
mass, this process is therefore less phase--space suppressed than the
$2\rightarrow4$ QCD (and EW) contributions. We found that for a chargino with
$\Delta m=m_{\charg}-m_{\st}=\gev{20}$, EW contributions are comparable to
the QCD contributions.
\begin{figure}[h]
\includegraphics[width=0.3\textwidth]{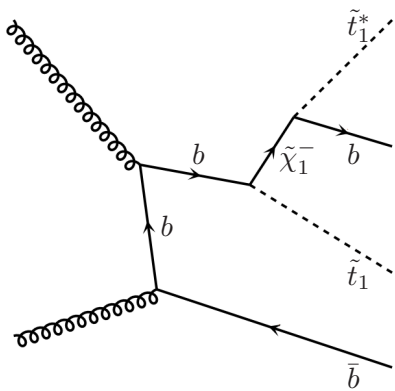}
\caption{Example diagram for EW stop pair production in association with two
  $b-$jets via gluon fusion. The chargino, $\tilde{\chi}_1^-$, might be
  on--shell.}
\label{fig:ew_t1t1bb}
\end{figure}

As motivated in the introduction we assume that $\tilde t_1$ is dominantly
right--handed. Hence, EW contributions are maximal if the light chargino is
higgsino--like. If it is wino--like EW contributions will be suppressed by
small mixing angles, because the wino does not directly couple to the
right--handed stop. Therefore, our results will not change significantly if
one adds light winos to our scenarios as long as two--body decays of the stop
and the higgsino into winos are kinematically forbidden.

As can be seen from Fig.~\ref{fig:ew_t1t1bb}, we are sensitive to the
$\charg-b-\st$ coupling. We will present prospects of determining the
$\charg-b-\st$ coupling in a later publication. If a signal is observed,
one should also be able to obtain information about the mass spectrum of the
chargino and stop sector. However, here we are primarily interested in the
discovery reach of this new $\tilde t_1$ search channel.

\section{LHC Analysis}
\label{sec:lhc_analysis}

In this section we first review the dominant SM background processes. Next we
choose a benchmark scenario in the co--annihilation region, which is
compatible with electroweak baryogenesis (if one adds an additional
CP--phase). We then present kinematic distributions and discuss our basic cuts
and compare the size of the EW contributions to the pure QCD prediction.
Finally, we show the discovery reach at the LHC in the neutralino stop mass
plane.

\subsection{Backgrounds}\label{subsec:backgrounds}

We only consider SM backgrounds, since we assume that all other colored
sparticles are quite heavy so that SUSY backgrounds are negligible. We look
for SM processes which lead to two $b-$jets and large missing energy. The
dominant backgrounds are 
\bit
\item $t\bar t$ production (including all top decay channels). Top decays will
  nearly always produce two $b-$jets. Since we require large missing $E_T$, at
  least one of the $W$ bosons produced in top decay will have to decay
  leptonically. Note that this also gives rise to a charged lepton ($e, \mu$
  or $\tau$), whereas the signal does not contain isolated charged leptons.

\item $Z(\rightarrow \nu \nu)+b \bar b$ production, {\it i.e.} $Z$ boson
  production in association with two $b-$jets. The $Z$ boson decays into a
  pair of neutrinos. If the charm jets in the signal are very soft, this
  background looks very similar to our signal. Fortunately it can be directly
  extracted from data. One can measure $Z(\rightarrow e^+e^-/\mu^+\mu^-)+b
  \bar b$, where the $Z$ decays into either a pair of electrons or muons.
  From the known $Z$ branching ratios (BRs) one can then obtain an estimate
  for the background cross section.  However, this procedure will increase the
  statistical error due to a smaller BR of the $Z$ to charged leptons compared
  to the decay into neutrinos \cite{Vacavant}.
\item $W(\rightarrow \ell \nu) + b \bar b$ production, where the $W$ decays
  leptonically. Again, this background will contain a charged lepton, and will
  thus resemble the signal only if the charged lepton is not identified. This
  can happen when the charged lepton emerges too close to the beam pipe or
  close to a jet; moreover, identification of hadronically decaying $\tau$
  leptons is not easy.
\item Single top production in association with a $b-$quark, e.g. $g u
  \rightarrow t \bar b d$. The second $b-$jet stems from top decay, and the
  missing $E_T$ comes from the leptonic decay of the $W$ boson.
\eit

We neglect QCD dijet and trijet production in our analysis, since a large
$\met$ cut should remove those backgrounds \cite{Aad:2009wy, atlasConf,
  Allanach:2010pp}.

Estimates for the total hadronic cross sections for these SM backgrounds are
given in Table~\ref{tab:bkg_xs}. The cross section for the $t \bar t$
background has been taken from Ref.\cite{Bonciani:1998vc}, which includes NLO
corrections as well as resummation of next--to--leading threshold
logarithms. All other backgrounds, as well as the signal, have been calculated
to leading order using {\tt Madgraph4.4.32} \cite{Maltoni:2002qb}.

\begin{table}[t]
\begin{center}
\begin{ruledtabular}
\begin{tabular}{c|cccc}
process & $Wb\bar b$ & single top & $Zb \bar b$ & $t\bar t$\\ \hline
$\sigma$ [pb] & 84 & 170 & 174 & 800\\
\end{tabular}
\end{ruledtabular}
\caption{Total hadronic cross sections in pb for the main SM backgrounds at
  $\sqrt{s}=14$ TeV. The cross sections were calculated with {\tt
    Madgraph} apart from $t\bar t$ production, which is calculated in
  Ref.~\cite{Bonciani:1998vc}. 
\label{tab:bkg_xs}}
\end{center}
\end{table}

We have generated $5\times10^{6}$ $t\bar t$, single top and $Wb\bar b$
background events, respectively, as well as $3\times10^{6}$ $Zb\bar b$ events.

\subsection{Numerical tools}

The masses, couplings and branching ratios of the relevant sparticles are
calculated with {\tt SPheno2.2.3} \cite{Porod:2003um}. We use the CTEQ6L1
parton distribution functions and the one--loop expression for the strong
gauge coupling with five active flavors with $\Lambda_{\rm{QCD}}=\mev{165}$
\cite{Pumplin:2002vw}. Our parton--level signal and background processes apart
from $t\bar t$ production are generated with {\tt Madgraph4.4.32}
\cite{Maltoni:2002qb}. Parton--level events are then interfaced with {\tt
Herwig++2.4.2} \cite{Bahr:2008pv} for the hadron--level simulation. We also
generate our $t\bar t$ events with {\tt Herwig++}, fixing the normalization as
in Ref.\cite{Bonciani:1998vc}. We do not consider detector effects. Jets are
reconstructed with {\tt FastJet2.4.1} \cite{Cacciari:2006sm} via the $k_t$
clustering algorithm with $R=1.0$. Our event samples are then analyzed with
{\tt HepMC2.04.02} \cite{Dobbs:2001ck} and {\tt ROOT} \cite{Brun:1997pa}. For
the sake of simplicity, we keep the $b-$ and $c-$flavored hadrons stable. A
jet is identified as a $b-$jet, if a stable $b-$hadron is found in the
reconstructed jet. If not otherwise mentioned, we assume a $b-$tagging
efficiency of $60\%$.

\subsection{Benchmark Scenario}
\label{sec:benchmark_sceanrio}

In order to develop a set of cuts, we introduce a benchmark scenario. We work
in the framework of the general MSSM with a light (dominantly right--handed)
stop with $m_{\st}=\gev{120}$, which is compatible with electroweak
baryogenesis. Our benchmark point is also consistent with DM constraints.  The
lightest neutralino is bino--like with
\beq
m_{\neutr}=m_{\st}-\gev{20}.
\label{eq:neut_mass_relation}
\eeq
The lightest chargino is higgsino--like with
\beq
m_{\charg}=m_{\st}+\gev{20}.
\label{eq:char_mass_relation}
\eeq

All other sparticles are decoupled. The cross section for our benchmark point
is given in the first line of Table~\ref{tab:signal_xs}. For comparison, we
separately present the total hadronic cross section for the QCD+EW
contributions (second column) and the pure QCD contribution (third column) as
well as the ratio of the respective cross sections (fourth column). We also
display the cross sections for heavier stops (first column) assuming the mass
relation of Eq.~(\ref{eq:char_mass_relation}).

We can see that the cross section decreases quickly with increasing sparticle
masses as expected. For example, increasing the stop mass from $\gev{120}$ to
$\gev{200}$ decreases the total cross sections by roughly a factor of
ten. However, as we will show in Sect.~\ref{sec:discovery_potential}, the
significance with respect to the SM backgrounds will decrease less
rapidly. This is because the final state particles will have on average larger
momenta and therefore the events will more easily pass our cuts. We also
observe that the EW contributions are significant and even give the leading
contribution to the cross section. For the case at hand, they enhance the
total hadronic cross section by roughly $150\%$ compared to the pure QCD
contribution.

Eq.~(\ref{eq:char_mass_relation}) implies that the decay $\st\rightarrow
\charg b$ is not allowed. Similarly, Eq.~(\ref{eq:neut_mass_relation}) implies
that $\st \rightarrow t \neutr$ and $\st\rightarrow b\neutr W$ decays are
forbidden. As mentioned in the introduction, our stop then decays via a flavor
changing neutral current (FCNC) decay, $\mathcal{BR}(\st\rightarrow c
\neutr)\approx1$. This requires that the physical $\tilde t_1$ has a
nonvanishing $\tilde c$ component. Even if squark flavor mixing is assumed to
be absent at some (high) input scale, it will be induced by electroweak
one--loop diagrams, due to the fact that quark generations do mix. This
one--loop process is enhanced by large logarithms \cite{Hikasa:1987db},
i.e. it can be understood as describing the running off--diagonal $\tilde t_L
\tilde c_L$ mass. Of course, it is also possible that squark mass matrices are
not exactly flavor diagonal at any scale. The stop generally decays promptly,
i.e. its flight path is much too short to be seen experimentally. However,
depending on the size of the $\tilde t_1 c \neutr$ coupling, the lifetime of
the stop can exceed $1/\Lambda_{\rm{QCD}}\approx10^{-24}\rm{s}$ in which case
the stop hadronizes before it decays \cite{Hiller:2009ii}. We assume that
$\mathcal{BR}(\charg\rightarrow \st b) = 1$; this almost follows from our
assumption that both (third generation) charged and neutral sleptons 
are heavier than $m_{\tilde t_1} - m_b$. 

\begin{table}
\begin{center}
\begin{ruledtabular}
\begin{tabular}{c|cc|c}
$m_{\st}$ [GeV] & $\sigma|_\text{QCD+EW}$ [pb] & $\sigma|_\text{QCD}$ [pb] &
  $\frac{\sigma|_{QCD+EW}}{\sigma|_\text{QCD}}$ \\ 
 \hline
 120 & 19 & 7.5 & 2.5 \\
 140 & 9.8 & 3.9 & 2.5 \\
 160 & 5.5 & 2.2 & 2.5 \\
 180 & 3.2 & 1.3 & 2.5 \\
 200 & 2.0 & 0.81 & 2.5 \\
 220 & 1.2 & 0.51 & 2.4 \\
 240 & 0.83 & 0.34 & 2.4 \\
 260 & 0.56 & 0.23 & 2.4 \\
 280 & 0.38 & 0.16 & 2.4 \\
 300 & 0.27 & 0.11 & 2.5 \\
 320 & 0.19 & 0.081 & 2.3 \\
\end{tabular}
\end{ruledtabular}
\caption{Total hadronic signal cross sections in pb from the pure QCD diagrams
  (third column) and from the QCD+EW contributions (second column) as a
  function of the stop mass (first column). The mass relation of
  Eq.~(\ref{eq:char_mass_relation}) has been assumed to hold. We also show in
  the fourth column the ratio of the QCD and QCD+EW cross sections. All cross
  sections were calculated with {\tt Madgraph} for $\sqrt{s}=14$ TeV. See
  Sect.~\ref{sec:signal_description} for further details.
\label{tab:signal_xs}}
\end{center}
\end{table}

\subsection{Distributions}

We present in this section kinematic distributions at the LHC of the
background and signal for our benchmark scenario. We show cumulative
distributions, i.e. the expected signal and the different background
contributions are stacked on top of each other. Note that we show the number
of events on a logarithmic scale. All distributions are scaled to an
integrated luminosity of $1\,\rm{fb}^{-1}$ at $\sqrt{s}=\tev{14}$. In all
cases we require at least two $b-$jets with a rapidity of
$|\eta_{b_{1,2}}|<2.5$. We also require the jets to have transverse momenta
$p_T>\gev{20}$. We assume in this subsection a $b-$tagging efficiency of one,
but we will assume later (in Sect.~\ref{sec:discovery_potential}) a more
realistic efficiency of $60\%$ when we derive the discovery potential. No
further cuts are applied.

In Fig.~\ref{fig:no_leptons}, we present the number of isolated charged
leptons (electrons, muons) for signal and background. We only consider
isolated leptons with $p_T^{\rm{lepton}}>\gev{5}$. A lepton is isolated, if
less than 10 GeV of energy are deposited in a cone of $\Delta
R=\sqrt{\Delta\phi^2+\Delta \eta^2}<0.2$ around the direction of the lepton
(not counting the energy of the lepton itself). The signal process possesses
nearly no isolated leptons, because no leptons can arise at parton level. We
thus employ a veto on isolated leptons in
Sect.~\ref{sec:discovery_potential}. This cut will effectively reduce the SM
backgrounds involving leptonically decaying $W$ bosons. Recall that we have
only considered the $Z\rightarrow \nu\bar \nu$ channel for the $Zb\bar b$
background. Since our $b-$hadrons are stable, we have no isolated leptons for
the $Zb\bar b$ background and at most two isolated leptons for the top
backgrounds. Later we will impose quite stiff cuts on the transverse momenta of
both $b-$jets; leptons originating from semileptonic $b$ decays would
therefore not be isolated. Similarly, the leptons resulting from semileptonic
charm decays will be either very soft or not isolated. Recall also that we
require the $W$ boson in the $W b \bar b$ background to decay leptonically,
while we do not demand specific decay modes for the top quarks in the single
top and $t \bar t$ backgrounds. However, these backgrounds can only produce
large missing $p_T$ if at least one $W$ boson decays leptonically. In that
case the result for the single top background would be very similar to that of
the $W b \bar b$ background, while the distribution for the $t \bar t$
background would peak at $n_\ell = 1$.

\vspace{+1.5mm}
\begin{figure}[t]
\vspace{+1.5mm}
\includegraphics[width=0.45\textwidth]{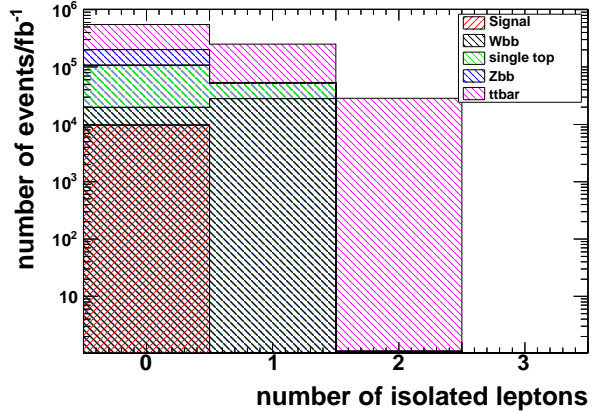}
\caption{Number of isolated leptons for the signal and SM backgrounds
assuming an integrated luminosity of $1\,\text{fb}^{-1}$ at $\sqrt{s}=14$ TeV.
For the signal we assumed the benchmark scenario of
Sect.~\ref{sec:benchmark_sceanrio}, {\it i.e.} $m_{\tilde{\chi}_1^0} =
\gev{100}$, $m_{\st} = \gev{120}$ and $m_{\tilde{\chi}_1^+} = \gev{140}$. 
The distributions are stacked on top of each other.}
\label{fig:no_leptons}
\end{figure}

Fig.~\ref{fig:bj1pt} shows the $p_T$ distribution of the hardest $b-$jet.  The
$t\bar t$ and single top backgrounds give the hardest $b-$jets, because here
at least one $b-$quark arises from the decay of a heavy top quark. The other
backgrounds have softer $b-$jets, which originate mainly from a $b\bar b$
pair generated via gluon splitting. For the signal, the dominant QCD
contribution is $\st\st^*$ production with an additional $b\bar b$ pair from
gluon splitting, whereas the main EW contributions are from $\st\charg b$
production, where the second $b$--quark comes from a resonant $\charg$ decay;
{\it cf.} Sect.~\ref{sec:signal_description}.

\begin{figure}[t]
\vspace{+1.5mm}
\includegraphics[width=0.45\textwidth]{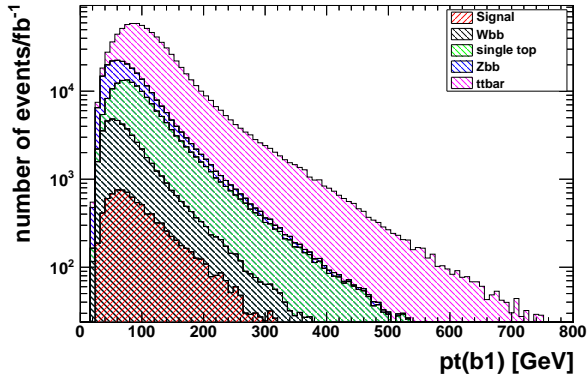}
\caption{Same as Fig.~\ref{fig:no_leptons}, but now for the $p_T$ distribution 
of the hardest $b$--jet.}
\label{fig:bj1pt}
\end{figure}

Note that for our benchmark scenario, in the mixed QCD+EW contribution the
$b-$quark from the $\charg$ decay is usually much softer than the other
$b-$quark. Moreover, $g \rightarrow b \bar b$ splitting prefers asymmetric
configurations, with one $b$ (anti)quark being significantly more energetic
than the other \footnote{The corresponding splitting function is proportional
  to $x^2 + (1-x)^2$, i.e. it peaks at $x=0$ and $x=1$.}. As a result the $p_T$
distribution of the second hardest $b-$jet shown in Fig.~\ref{fig:bj2pt} is
much softer for the signal, peaking around $\gev{40}$, whereas the $p_T$
distribution of the hardest signal $b-$jet peaks around $\gev{70}$; see
Fig.~\ref{fig:bj1pt}. The distribution of the second $b-$jet in the single top
background is also very soft, since it originates from $g \rightarrow b \bar
b$ splitting in the initial state, and that for the $W b \bar b$ background is
soft since here the $b-$jets originate from $g \rightarrow b \bar b$ splitting
in the final state. In contrast, the main contribution to the $Z b \bar b$
background can be understood as $gg \rightarrow b \bar b$ production where the
$Z$ boson is emitted off the quark line; here, and in the $t \bar t$
background, the difference between the $p_T$ distributions of the two $b-$jets
is therefore smaller than for the other processes.

The reason for the soft $p_T$ spectrum of the second signal $b-$jet is the
small mass splitting of $\gev{20}$ between the $\st$ and the $\charg$. In the
rest frame of the decaying chargino, the $b-$quark has 3--momentum $|\vec
p^*|=\gev{10}$. If we increase the mass gap between stop and chargino, the
$p_T(b_2)$ distribution will be much harder, {\it e.g.} for $m_{\charg} =
\gev{210}$ we obtain $|\vec p^*|=\gev{70}$. However, increasing $m_{\tilde
  \chi_1^\pm}$ also decreases the signal cross section significantly. For
example, increasing the stop chargino mass difference from $\gev{20}$ to
$\gev{80}$ reduces the total hadronic cross section from $19\,\rm{pb}$ (see
Table~\ref{tab:signal_xs}) to $10\,\rm{pb}$. In this case 75\% of the cross
section would come from pure QCD contributions. This reduction of the cross
section overcompensates the gain of efficiency due to the harder $p_T(b_2)$
spectrum, i.e. the signal cross section after cuts also decreases with
increasing chargino mass, although less quickly than the total cross section
before cuts does.

\begin{figure}[t]
\vspace{+1.5mm}
\includegraphics[width=0.45\textwidth]{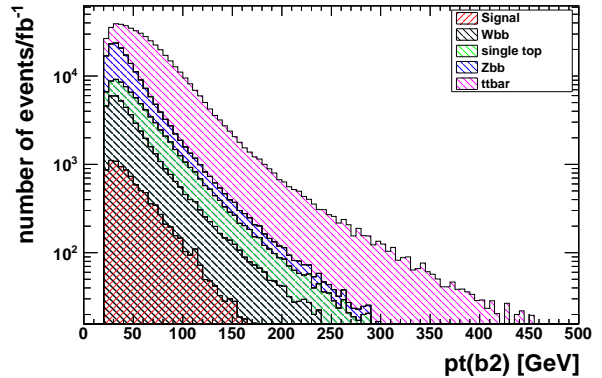}
\caption{Same as Fig.~\ref{fig:no_leptons}, but now for the $p_T$ distribution 
of the second hardest $b$--jet.}
\label{fig:bj2pt}
\end{figure}

With the help of Figs.~\ref{fig:bj1pt} and \ref{fig:bj2pt}, we found lower
cuts of $\gev{150}$ ($\gev{50}$) on the hardest (second hardest) $b-$jet
helpful in order to increase the signal to background ratio. We will employ
these cuts in the next subsection. A similar cut on the hardest jet (and on
the missing energy) is also required by the trigger \cite{trigger}.

The missing transverse momentum, $\mpt$ distributions of the signal and the
backgrounds are given in Fig.~\ref{fig:ptmiss}. We observe that the signal
$\mpt$ distribution falls off less quickly than the background
distributions. Therefore, we will employ a lower cut on $\mpt$ of
$\gev{200}$. On the one hand, this cut increases the significance of the
signal over the top, $Zb \bar b$ and $Wb \bar b$ backgrounds. On the other
hand, we expect that it suppresses pure QCD backgrounds like dijet and trijet
production to a negligible level \cite{Allanach:2010pp,Aad:2009wy}.

\begin{figure}[t]
\vspace{+1.5mm}
\includegraphics[width=0.45\textwidth]{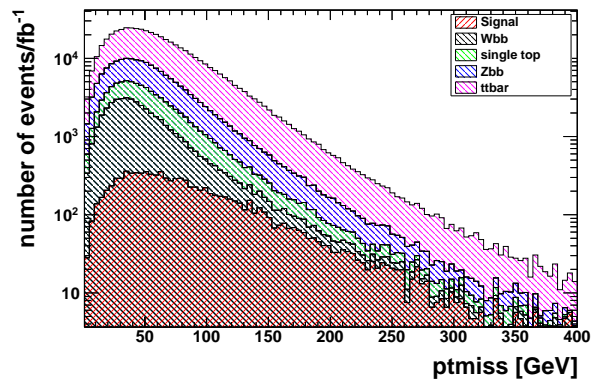}
\caption{Same as Fig.~\ref{fig:no_leptons}, but now for the missing 
transverse momentum distribution.}
\label{fig:ptmiss}
\end{figure}
 
In addition to kinematic distributions we can also employ the number of
charged particles (mainly hadrons) to distinguish signal from background
events. The respective distributions are given in Fig.~\ref{fig:charge}, where
only charged particles with $p_T>\gev{2}$ and $|\eta|<2.5$ are
included. Moreover, we have assumed the $W$ boson in single top events as well
as at least one $W$ boson in $t \bar t$ events to decay leptonically 
in order to obtain a significant amount of missing energy
\footnote{Figs.~\ref{fig:charge} and \ref{fig:ratio} show only 
single top and $t \bar t$ events with at least one isolated electron 
or muon in the final state or events with at least one hadronically decaying tau.
We employed the lepton cuts of Sect.~\ref{sec:discovery_potential}.}.  The
number of charged particles, $N_{\rm charged}$, in $t \bar t$ events is
nevertheless on average larger than for the signal. The second $W$ boson
will usually decay hadronically, producing jets that are usually considerably
harder than the $c-$jets in the signal.  Recall that single top production is
dominated by $g u \rightarrow t \bar b d$, giving a $W b \bar b d$ final state
after top decay. The resulting multiplicity distribution looks similar to that
of the signal. In contrast, the averaged charged particle multiplicity for the
$Wb\bar b$ and $Zb\bar b$ backgrounds is slightly lower than for the
signal. This is because the charm quarks from stop decay generate on average
more charged particles than the gauge boson decay products, which contain
usually only one or three charged particles (the latter being due to 3--prong
$\tau$ decays).  We have tried several cuts for the charged particle
multiplicity. We will choose $N_{\rm charged} \geq 10$ in
Sect.~\ref{sec:discovery_potential}, which reduces the $Wb\bar b$ and $Zb \bar
b$ backgrounds relative to the signal.

\begin{figure}[t]
\vspace{+1.5mm}
\includegraphics[width=0.45\textwidth]{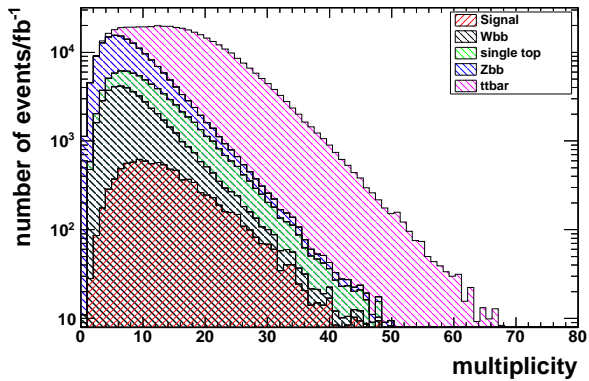}
\caption{Same as Fig.~\ref{fig:no_leptons}, but now for the
multiplicity of charged particles with $|\eta|<2.5$ and $p_T>\gev{2}$;
moreover, we have required that single top and $t \bar t$ events contain at least
one leptonically decaying $W$ boson.}
\label{fig:charge}
\end{figure}

Finally, Fig.~\ref{fig:ratio} shows the number of events as a function of the
ratio between the $p_T$ of the hardest $b-$jet and the $\mpt$. The signal
distribution has a steeper fall--off than those for the background processes
and roughly peaks at one. This is not unexpected, because $p_T(b_1)$ and
$\mpt$ are strongly correlated for the signal, due to the relative softness of
the second $b-$jet and the $c-$jets. This means that the $\tilde t_1 \tilde
t_1^*$ pair, whose $p_T$ roughly corresponds to the missing $p_T$, essentially
recoils against the harder $b-$jet. From this simplified picture we would expect a
ratio of $\lesssim 1$. In contrast, the single top and $t \bar t$ backgrounds
contain more partons, weakening the correlation between the missing $p_T$ and
that of any one jet.

\begin{figure}[t]
\vspace{+1.5mm}
\includegraphics[width=0.45\textwidth]{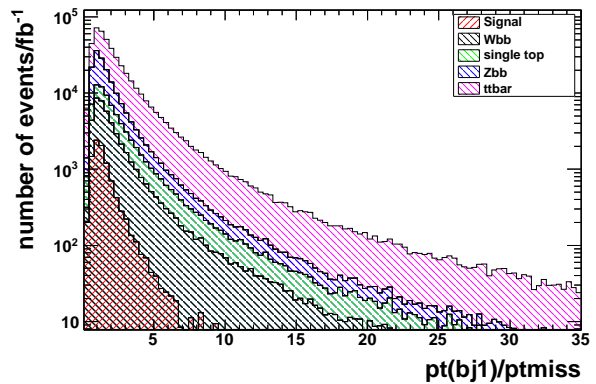}
\caption{Same as Fig.~\ref{fig:no_leptons}, but now for the ratio of the $p_T$
  of hardest $b$--jet to the missing transverse momentum; moreover, we have
  required that single top and $t \bar t$ events contain at least one leptonically
  decaying $W$ boson.}
\label{fig:ratio}
\end{figure}

\subsection{Discovery Potential at the LHC}
\label{sec:discovery_potential}

After we have discussed the basic cuts for our benchmark scenario, we now turn
to the discovery potential for our $b\bar b$ plus missing energy signature at
the LHC. We present numerical results for our benchmark scenario and for
heavier stop masses. We show the statistical significance for the pure QCD
prediction of the signal and the QCD+EW prediction as a function of the stop
mass.  We shortly discuss the effect of varying chargino masses on the
statistical significance.  We also present the statistical significance as a
function of the stop and neutralino mass.

From the discussion of kinematical distributions and particle multiplicities
in the previous subsection, we find that the following cuts maximize the
significance of our signal: 
\bit
\item $N_{\rm{b-jets}} \geq 2$, i.e. we require at least two tagged $b-$jets
  with $p_T>\gev{20}$ and $|\eta|<2.5$.
\item $N_{\rm{lepton}}<1$, i.e. we veto all events with an isolated electron or
  muon with $|\eta|<2.5$ and $p_T>\gev{5}$.
\item $p_T(b_1)>\gev{150}$, i.e. we require a large transverse momentum for
  one of the $b-$jets.
\item $p_T(b_2)>\gev{50}$, i.e. we impose a significantly weaker cut on the
  second $b-$jet.
\item $\mpt>\gev{200}$, i.e. we demand large missing transverse momentum.
\item $N_{\rm{charged}} \geq 10$, i.e. require at least ten charged particles
  with $p_T>\gev{2}$ and $|\eta|<2.5$.
\item $\frac{p_T(b_1)}{\mpt}<1.6$, i.e. we demand that the ratio of the $p_T$ of
  the most energetic $b-$jet and $\mpt$ is not large.
  \eit

Table~\ref{tab:cut_flow} shows the effect of these for our benchmark scenario,
assuming an integrated luminosity of $100\,\rm{fb}^{-1}$. Numbers are given
with a $b-$tagging efficiency of $60\%$. The statistical significance is
estimated with $\mathcal{S}=\frac{S}{\sqrt{B}}$, where $S$ and $B$ are the
number of signal events and background events, respectively. We also show the
ratio $S/B$.

We start Table~\ref{tab:cut_flow} with requiring two tagged $b-$jets with
$p_T>\gev{20}$, as well as $\mpt>\gev{200}$. With these cuts, we already
obtain a statistical significance of about $49$, but $S/B$ is around 0.1. 

After applying the cut on the $p_T$ of the hardest $b-$jet, the statistical
significance increases to $\mathcal{S}=51$, and the signal to background ratio
improves to about 1/8. Note that this rather stiff cut has relatively little
effect on the signal as well as on the $W b \bar b$ and $Z b \bar b$
backgrounds, since the even stiffer $\mpt$ cut implies a rather hard spectrum
for one of the $b-$jets in these cases. On the other hand, this cut does
reduce the $t \bar t$ background significantly. 

After the $p_T$ cut on the second $b-$jet, the statistical significance falls
to $\mathcal{S}=44$. However, for a lower $p_T$ cut on the second $b$--jet the
tagging efficiency worsens \cite{Aad:2009wy}, so we keep $p_T>\gev{50}$. Note
also that this cut slightly increases $S/B$.

\begin{table*}[t!]
\begin{center}
\begin{ruledtabular}
\begin{tabular}{c||c|c|c|c|c||c|c}
cut & $Wbb$ & single top & $Zbb$ & $t\bar t$  & Signal & $S/B$& $S/\sqrt{B}$ \\
\hline
$\mpt>\gev{200}$, 2 $b$--jets with $|\eta|\le 2.5$ &  $6\,144$ & $10\,390$ &
$33\,440$& $179\,900$ & $23\,360$ & 0.098 & 49\\ 
$p_T(b_1)>\gev{20}$, $p_T(b_2)>\gev{20}$ & & & & & & \\\hline
$p_T(b_1)>\gev{150}$ & $5\,765$ & $7\,824$ & $27\,720$ & $127\,200$ & $20\,760$
& 0.123 & 51 \\ \hline 
$p_T(b_2)>\gev{50}$ & $4\,269$ & $5\,476$ & $19\,290$ & $92\,330$ & $15\,360$
& 0.127 & 44 \\ \hline 
veto on isolated leptons & $1\,286$ & $2\,373$ & $19\,290$ & $32\,400$ &
$15\,360$ & 0.278 & 66\\\hline 
$\#$ charged hadrons $\geq$ 10 & $1\,096$ & $2\,227$ & $15\,570$ & $32\,200$ &
$15\,020$ & 0.293 & 67 \\ \hline 
$p_T(b_1)/\mpt<1.6$ & $881$ & $1\,485$ & $14\,970$ & $22\,030$ & $13\,700$
& 0.348 & 69\\ 
\end{tabular}
\end{ruledtabular}
\caption{Cut flow for the benchmark scenario of
  Sect.~\ref{sec:benchmark_sceanrio} at the LHC with $\sqrt{s}=14$ TeV and an
  integrated luminosity of $100\,\rm{fb}^{-1}$. Numbers are given for a
  $b$--tagging efficiency of $0.6$. 
\label{tab:cut_flow}}
\end{center}
\end{table*}

The lepton veto greatly reduces the SM backgrounds involving leptonic $W$
decays. Our statistical significance is now $\mathcal{S}=66$, with a signal to
background ratio better than 1/4. Most $W b \bar b$, $t \bar t$ and single top
background events passing this cut contain a hadronically decaying $\tau$
lepton, so a hadronic $\tau$ veto would suppress these backgrounds even
further. We will come back to this point shortly.

The cut on the charge multiplicity further suppresses the $W b \bar b$ and $Z
b \bar b$ backgrounds, increasing the statistical significance to
$\mathcal{S}=67$ and the signal to background ratio to nearly 0.3.

In contrast, the final cut, on the ratio $p_T(b1)/\mpt$, is quite efficient at
suppressing the single top and top pair backgrounds. For our benchmark
scenario we now obtain a statistical significance of $\mathcal{S}=69$, and a
signal to background ratio of about 0.35.  A good signal to background ratio
is important for a precise determination of the $\st-\charg-b$ coupling.
Furthermore, knowledge of the systematic error of the SM backgrounds at the
10$\%$ is sufficient for our benchmark point to observe the signal. With an
integrated luminosity of $100\,{\rm fb}^{-1}$, such a precision is expected to be
easily obtained by the experimental groups \cite{Aad:2009wy}.

As mentioned above, after vetoing events containing an isolated electron or
muon, most SM background events originating from a leptonically decaying $W$
boson will have a $\tau-$jet in the final state. Requiring a $\tau$ veto
should reduce the SM background and therefore enhance the significance and the
signal to background ratio. Usually one is interested in identifying, rather
than vetoing, $\tau$ jets. The most common method is to look for jets above a
certain $p_T$ threshold that contain only one or three charged particles, and
not too much neutral energy; this greatly suppresses backgrounds from QCD
jets. As an example, assuming a $\tau$ tagging efficiency of $50\%$ for $\tau$
leptons with $p_{T,\tau} > 15$ GeV and $|\eta_\tau| < 2.5$ we find $7693$ background
events with at least one $\tau$ in the final state. At the same time, the
number of signal events stays nearly the same, if we assume a small mistagging
probability. Therefore, the significance is increased to $77$, while the
signal to background ratio increases to $0.43$. Vetoing $\tau-$jets does not
reduce the $Z b \bar b$ background; recall, however, that this can be
subtracted reliably using $Z$ decays into electrons or muons.

The details of the $\tau$ jet identification algorithm can be tuned, allowing
to increase the efficiency for correctly identifying $\tau-$jets at the cost
of increasing the rate with which QCD jets are misidentified. For example,
Ref.~\cite{tau_id} finds a 50\% $\tau$ tagging efficiency and a $\sim98\%$ QCD
jet rejection efficiency for $p_T(\tau_{\rm jet}) \lsim 28$ GeV. In the case
at hand it would probably be better to have a higher efficiency for
identifying $\tau-$jets, even at the cost of more false positives. Recall that
at least in the absence of ISR and FSR, additional jet activity in signal
events results from the typically rather soft $c-$quarks produced in $\tilde
t_1$ decay. The probability for misidentifying a $c-$jet as a $\tau-$jet
presumably differs from that for light quark or gluon jets. Detailed knowledge
of the detector is thus required for optimizing the $\tau$ identification
(or rather, veto) criteria in our case. We therefore do not pursue this avenue
further.

\begin{figure}[h]
\vspace{+1.5mm}
\includegraphics[width=0.52\textwidth]{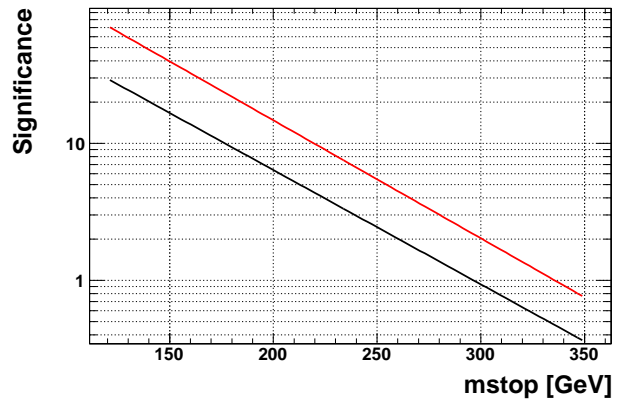}
\caption{Significance for QCD contributions only (black line) and QCD+EW (red
  line) as a function of the stop mass at the LHC with $\sqrt{s}=\tev{14}$. We
  assume an integrated luminosity of $100\,\rm{fb}^{-1}$ and the mass
  relations of Eq.~(\ref{eq:neut_mass_relation}) and
  Eq.(\ref{eq:char_mass_relation}). 
 } 
\label{fig:mstop_mchargconst}
\end{figure}

So far, we have discussed the significance for a light stop. Now, we turn to
larger stop masses. We apply the same mass relations for $\neutr$ and $\charg$
as for our benchmark scenario, {\it i.e.}  Eqs.~(\ref{eq:neut_mass_relation})
and (\ref{eq:char_mass_relation}). We use the cuts of Table~\ref{tab:cut_flow}. 
In Fig.~\ref{fig:mstop_mchargconst}, we
present the significance for the pure QCD case (black line) and the QCD+EW
contributions (red line) as a function of the stop mass. We assume again an
integrated luminosity of $100\,\rm{fb}^{-1}$ and a $b-$tagging efficiency of
$60\%$. The significance for scenarios with the same stop and neutralino mass,
but with heavier charginos, will lie between these two cases; in practice the
EW contribution will be negligible (for the given set of cuts) if $m_{\tilde
  \chi_1^\pm} \gsim 2 m_{\tilde t_1}$.

Evidently the EW contribution increases the significance by a factor $\gsim
2$; this is very similar to the increase of the total signal cross section
found in Table~\ref{tab:signal_xs}. After the cuts of Table~\ref{tab:cut_flow}
the statistical $5\sigma$ discovery reach extends to $\gev{260}$ ($\gev{210}$)
with (without) the EW contributions.

\begin{figure}[t]
\vspace{+1.5mm}
\includegraphics[width=0.45\textwidth]{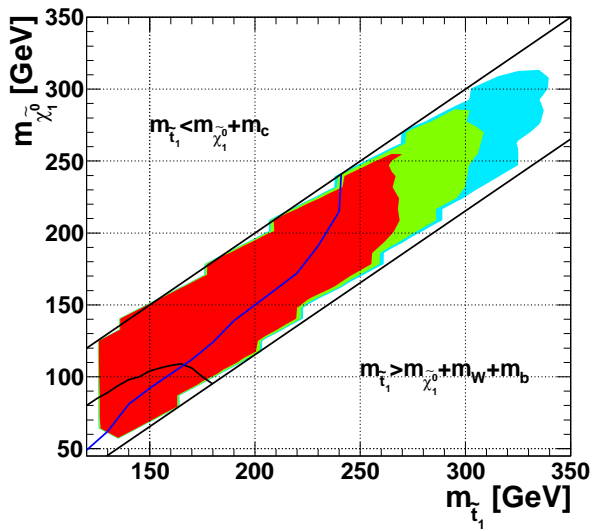}
\caption{Statistical signal significance at the LHC with $\sqrt{s}=14$ TeV as
  a function of the stop and neutralino mass. The red, green and turquoise
  region corresponds to an excess of at least 5$\sigma$, 3$\sigma$ and 2$\sigma$, respectively, 
  for an integrated luminosity of $100\,\rm{fb}^{-1}$. The chargino mass is fixed by
  Eq.~(\ref{eq:char_mass_relation}). The parameter space {\em below} the black
  curve is excluded by Tevatron searches \cite{CDFnote9834}, while in the
  region {\em above} the blue curve, the stop pair plus single jet
  (``monojet'') signal has significance $\geq 5$
  \cite{Carena:2008mj}. The parameter region where $\tilde t_1$ decays
  into a charm and a neutralino are expected to dominate is given by the
  condition $m_{\tilde{\chi}_1^0}+m_c < m_{\tilde{t}_1} < m_{\tilde{\chi}_1^0}
  + m_W + m_b$. }
\label{fig:mstop_mneut_mchargconst}
\end{figure}

In Figure~\ref{fig:mstop_mneut_mchargconst}, we show the significance in the
stop--neutralino mass plane for an integrated luminosity of
$100\,\rm{fb}^{-1}$ at $\sqrt{s}=14$ TeV. We assume the mass relation of
Eq.~(\ref{eq:char_mass_relation}). The parameter space {\em below} the black
curve is excluded by Tevatron searches at the 95\% confidence level
\cite{CDFnote9834}, whereas the parameter points {\em above} the blue curve
allow detection of a monojet signal with at least $5\sigma$ significance  
\cite{Carena:2008mj}. In our analysis, we assume that the light
stop decays dominantly into a charm quark and the lightest neutralino. This
will generally not be the the case if the mass relations
$m_{\tilde{t}_1}>m_{\tilde{\chi}_1^0}+m_c$ and $m_{\tilde{t}_1} <
m_{\tilde{\chi}_1^0} + m_W + m_b$ do not hold.

We first observe in Figure~\ref{fig:mstop_mneut_mchargconst} that the
significance increases with decreasing stop--neutralino mass
difference \footnote{Small deviations from this behaviour
are due to statistical fluctuations of the signal.}.
Recall that we require $p_T(b_1) > \gev{150}$. This implies that
the $\tilde t_1 \tilde t_1^*$ pair usually has a rather large transverse
momentum, which in turn leads to significant boosts of the stop squarks. As a
result the charm jets tend to go into a direction close to that of the stop
pair; i.e. the charm jets tend to reduce the missing $p_T$. Scenarios with
softer charm jets, i.e. smaller $\st-\neutr$ mass splitting, therefore pass
the $\mpt$ cut more efficiently. A similar argument also holds for the
significance of the monojet signal, where the missing energy is
is required to have ${\not p_T} > \tev{1}$; moreover, a veto against additional
jets is imposed \cite{Carena:2008mj}.

Our results suggest that a discovery of stops via stop pair production in
association with two $b-$jets is possible as long as $m_{\tilde{t}_1} \lesssim
\gev{270}$. In the case of non-observation of any signal one would be able to
exclude the parameter space at $2\sigma$ for stop masses of up to
$\gev{340}$; {\it cf.} the turquoise region in Fig.~\ref{fig:mstop_mneut_mchargconst}. 
Our signal is also visible in regions of parameter space where
the monojet signature produces no significant excess over the SM backgrounds
\footnote{We note however that in contrast to our work, Ref.~\cite{Carena:2008mj} 
included systematic uncertainties equal to $\sqrt{7B}$; see Ref.~\cite{Carena:2008mj}
for more details. The monojet discovery reach will therefore increase
if one uses solely $S/\sqrt{B}$ as an estimator for the significance.}. This is
partly because our process is additionally enhanced by the EW contributions,
which do not enter significantly the monojet production process.  We note
however, that different cuts for the monojet signal than those of
Ref.~\cite{Carena:2008mj} might increase the monojet discovery
reach \footnote{We did some estimates for the monojet signal and the most
  important backgrounds and found that less hard cuts on the monojet $p_T$ and
  on the ${\not p_T}$ lead in general to larger significancies. We plan to
  present our results in a future publication.}.

\section{Possibilities to further optimize the cuts}\label{sec:charm_tagging}

So far we have only applied cuts that can be used for all combinations of stop
and LSP masses, i.e. we have not tried to optimize the cuts as function of
$m_{\tilde t_1}$ or $m_{\tilde \chi_1^0}$. This simplifies the statistical
analysis: if we try several combinations of cuts, the chance of finding an
upward fluctuation of the background increases. On the other hand, additional
cuts could help to increase not only the statistical significance, but also
the purity of the sample, i.e. the signal to background ratio. Note that
Table~\ref{tab:cut_flow} implies a signal to noise ratio of about 0.025 at the
limit where the signal becomes statistically significant, i.e. for $S/
\sqrt{B} = 5$. If we require $S/B > 0.1$, the search reach would decrease from
about 270 GeV to 190 GeV. After imposing the $\tau$ veto, this would improve
again to about 200 GeV, which is however still much smaller than the
statistical reach.

Recall that our background is dominated by $Z b \bar b$ and $t \bar t$
events. The systematic uncertainty on the former is very small, since one can
analyze events with identical kinematics where the $Z$ decays into an $e^+e^-$
or $\mu^+\mu^-$ pair. Of course, $t \bar t$ events will also be analyzed in
great detail by the LHC experiments; nevertheless some extrapolation into the
signal region will presumably be required to determine this background, which
will introduce some systematic uncertainty.

Note also that the statistical $5\sigma$ discovery limit after the cuts of
Table~\ref{tab:cut_flow} still corresponds to about $1\,000$ signal
events. Refined cuts could therefore also help to further increase the mass
reach. 

In this section we therefore first discuss some cuts that might help to
suppress the $t \bar t$ background for scenarios with small $\st - \neutr$
mass splitting. We then discuss possibilities to identify the $c-$jets in the
signal, which can be used to suppress all backgrounds, and also to distinguish
our scenario from other SUSY processes with a similar final state.

\subsection{Further cuts for small mass splitting}

If $m_{\tilde t_1} - m_{\tilde \chi_1^0}$ is small, the $c-$jets will
usually be too soft to be detected. In contrast, many $t \bar t$ background
events will contain a hadronically decaying $W$ boson in addition to the two
required $b-$jets. We saw in Fig.~\ref{fig:charge} that this leads to a higher
charged multiplicity of these background events. One can cut on several
(related) quantities to suppress this background (and, to a lesser extent, the
single top background):
\bit
\item A jet veto could be imposed. Recall, however, that the signal is almost
  exclusively due to gluon fusion, which implies strong initial state
  radiation. In order not to unduly suppress the signal it might therefore be
  preferable to only veto events that contain two (or more) additional hard jets.
\item An upper limit on the total transverse energy could be imposed; probably
  the two $b-$jets should not be included here, i.e. one should only sum over
  particles (or calorimeter cells) that are not part of these two jets.
\item An upper limit on the charged particle multiplicity could be
  imposed. For example, requiring $N_{\rm charged} \leq 20$ would reduce the $t
  \bar t$ background by $\sim 30\%$ with little loss of signal.
\eit

\subsection{Charm tagging}

Observing two soft charm jets in addition to two hard $b-$jets and large
missing energy would give strong evidence for our scenario. However, soft
charm jets are probably difficult to identify at the LHC. The underlying
event, QCD radiation and possibly overlapping events (which we did not include
in our simulation) generate substantial hadronic activity in our signal events
and may overshadow potential charm--jets. That said, we can still expect some
reasonably energetic $c-$jets for the light stop scenarios if the $\st -
\neutr$ mass splitting is not too small. For example, for our benchmark point
and after applying all the cuts of Table~\ref{tab:cut_flow} we expect about
$45\%$ of all signal events to contain at least one $c-$jet with $p_T >
\gev{50}$ \footnote{Recall that we use $R=1$ in our jet definition. This
  allows a significant contribution from the underlying event and from
  unrelated QCD radiation to the jet energy. In the absence of these effects,
  only $\sim 16\%$ of all signal events contain a $c-$jet with $p_T \geq
  \gev{50}$.}.

Still, we have to discriminate the charm jets from light--flavored and gluon
jets. In Ref.~\cite{Carena:2008mj}, the authors use the jet mass and charged
particle multiplicity inside the jet to discriminate between $c-$jets and
light flavor or gluon jets. They find $c-$tagging efficiencies of $\geq 50\%$
for mass splitting $\geq \gev{10}$, but with a 25\% mis--tagging
probability. In the case at hand it might be better to employ a more
sophisticated algorithm, e.g. based on a neural net, that attempts to decide
whether the hadronic activity not associated with the two $b-$jets is
consistent with containing two $c-$jets, rather than identifying the
$c-$jets in isolation.

Identifying at least one $c-$jet would suppress the $W b \bar b$, single top
and $Z b \bar b$ backgrounds. Moreover, it would distinguish the stop pair
production we are considering from the pair production of light sbottom
squarks ($\tilde b_1$), followed by $\tilde b_1 \rightarrow b \neutr$ decays,
which also leads to a final state with two $b-$jets and large missing $p_T$,
but without $c-$jets. However, half of all $t \bar t$ events containing a
hadronically decaying $W$ boson also contain a $c-$jet (from a $W^+
\rightarrow c \bar s$ decay or its charge conjugate); $c-$tagging will
therefore probably not be as efficient in reducing the $t \bar t$ background.

\section{Summary and Conclusion}\label{sec:summary}

In this article, we analyzed stop pair production in association with two
$b-$quarks in the co--annihilation region, where the $\st - \neutr$ mass
difference is small, so that simple stop pair production does not yield an
observable signal.  Assuming a light higgsino--like chargino beside a stop
NLSP and neutralino LSP, the leading order QCD as well as mixed QCD--EW
contributions are taken into account. We discussed in some detail the dominant
diagrams for our signal process. We simulated the signal process and dominant
SM background processes. For a benchmark scenario with $m_{\st}=\gev{120}$,
$m_{\neutr}=\gev{100}$ and $m_{\charg}=\gev{140}$, we described all important
kinematic distributions and developed a set of cuts that help to isolate our
signal process. The most effective cut is a veto on isolated leptons. We
showed that we can have a significance of $\mathcal{S}=70$ for our benchmark
point, assuming an integrated luminosity of $100\,{\rm fb}^{-1}$ at $\sqrt{s} =
\tev{14}$; alternatively, a signal with $5\sigma$ statistical significance can
be obtained with just $0.5 \, {\rm fb}^{-1}$ of data. For fixed $\charg-\st$
and $\st-\neutr$ mass differences, the same set of cuts allows to detect the
signal at $\geq5 \sigma$ statistical significance for $m_{\st} \lesssim
\gev{270}$. Note that the significance of our signal actually increases when
the $\st - \neutr$ mass difference is reduced, making it complementary to
searches for direct stop pair production. The inclusion of the EW diagrams
greatly enhances the total cross section and more than doubles the statistical
significance.

However, we saw that for $m_{\st} > 190$ GeV our cuts leave us with a
relatively poor signal to background ratio. This can be improved by vetoing
$\tau-$jets, i.e. hadronically decaying $\tau$ leptons. We also briefly
discussed additional cuts on the hadronic activity not associated with the two
hard $b-$jets that could help to further suppress the $t \bar t$ background,
which probably will have larger systematic uncertainties than the $Z b \bar b$
background. These cuts should be tailored to the $\st-\neutr$ mass difference,
and perhaps also to the overall mass scale (which affects the amount of QCD
radiation); we have therefore not attempted to investigate this
systematically. We also briefly discussed the possibility to use charm tagging
to further suppress the backgrounds, and to distinguish our process from other
SUSY reactions yielding two $b-$jets and missing $p_T$. Charm tagging, and
$\tau$ vetoing, depend quite strongly on details of the detector performance;
a quantitative analysis of their efficiency is therefore best left to our
experimental colleagues.

One motivation for analyzing this final state is that the mixed QCD--EW
production channels might allow to probe the $\st-\charg-b$ coupling, thereby
allowing for the first time to check a SUSY coupling relation involving Yukawa
couplings; we intend to investigate the feasibility of determining this
coupling in a later publication. This will require that the $\st$ mass is
known. Determining both the mass and the EW coupling should be easier if the
production of stop pairs recoiling against a photon or a very hard jet
\cite{Carena:2008mj} also leads to an independent statistically significant
signal, which does not depend on the $\st-\charg-b$ coupling.


\begin{acknowledgments}
We thank Sebastian Fleischmann and Stefano Profumo for helpful discussions.
The work of J.S.K. is supported in part by the Initiative and Networking Fund
of the Helmholtz Association, contract HA-101 ("Physics at the
Terascale"). The work of of S.G. is supported in part by the U.S. Department
of Energy, under grant number DE-FG02-04ER41268 and in part by a Feodor Lynen
Research Fellowship sponsored by the Alexander von Humboldt Foundation. 
J.S.K. and S.G. thank the University of Bonn and the Bethe Center for
hospitality during numerous visits.
\end{acknowledgments}

\bibliographystyle{h-physrev}


\end{document}